\documentclass[aps]{revtex4}
\usepackage[colorlinks=true, pdfstartview=FitV, linkcolor=blue, citecolor=red, urlcolor=magenta, breaklinks=true, pdftex]{hyperref}
\usepackage{graphicx}  
\usepackage{amsfonts}
\begin{document}
\title{Quantum-corrected self-dual black hole entropy in tunneling formalism with GUP }
\author{M. A. Anacleto, F. A. Brito, E. Passos}
\email{anacleto, fabrito, passos@df.ufcg.edu.br}

\affiliation{Departamento de F\'{\i}sica, Universidade Federal de Campina Grande, Caixa Postal 10071, 58109-970 Campina Grande, Para\'{\i}ba, Brazil}

\newcommand{\m}{\mu}
\newcommand{\n}{\nu}
\newcommand{\s}{\sigma}

\def\eq#1{(\ref{#1})}

\begin{abstract}
In this paper we focus on the Hamilton-Jacobi method to determine the entropy of a self-dual black hole by using linear and quadratic GUPs(generalized uncertainty principles). 
We have obtained the Bekenstein-Hawking entropy of self-dual black holes and its quantum corrections that are logarithm and also of several other types. 
\end{abstract}

\maketitle

\section{Introduction}
The study of the black hole has been the subject of great interest from the fundamental physics in the last decades.
Black holes constitute an important class of exact solutions
of Einstein's equations which are characterized by mass ($M$), electric charge ($Q$) and angular momentum ($\vec{J}$)~\cite{Andrei:2005} and play a central role in both classical and quantum gravitational physics.
In particular a regular static black hole metric, known as loop black hole (LBH) or self-dual black hole, with quantum gravity corrections inspired by loop quantum gravity was derived in~\cite{Modesto:2009ve}.
Loop quantum gravity is based on a canonical quantization of the Einstein equations
written in terms of the Ashtekar variables~\cite{Ashtekar:1966}, that
is in terms of a $SU(2)$ 3-dimensional connection $A$ and a
triad $E$. 
The self-dual black hole has an event horizon and a Cauchy horizon --- see for instance~\cite{mhorizon,mh, ABParea} to the study of black hole which possess more than one horizon in order to understand the origin of black hole entropy in microscopic level.

A characteristic of self-dual black hole is self-duality property that removes the singularities
and replaces it with another asymptotically flat region. {We should mention that in supernormalizable gravity theories (an alternative to reconcile general relativity and quantum mechanics) there also appear spherically symmetric solutions as regular black hole solutions. For suitable choice of parameters, curvature tensors at the origin are finite. On the other hand, multi-horizon black holes can be also found depending on their masses \cite{Modesto:2011kw,Modesto:2013ioa}.}
The study of self-dual black hole and its thermodynamic property was analyzed in \cite{l.modesto-prd80, e.alesci-grqc-11015792, s.hossenfelder-grqc12020412}, and the dynamical aspects of the collapse 
and evaporation were studied in \cite{s.hossenfelder-prd81}.

A semiclassical approach considering the Hawking radiation as a tunneling phenomenon across the horizon has been proposed in recent years \cite{mk.parikh-prl85, mk.parikh-hepth0402166, ec.vagenas-plb559}.
In this approach the positive energy particle created just inside the horizon can tunnel through the geometric barrier quantum mechanically, and it is observed as the Hawking flux at infinity.
There are several approaches to obtain the Hawking radiation and the entropy of black holes. One of them is the 
Hamilton-Jacobi method which is based on the work of Padmanabhan and collaborators~\cite{SP} and also the effects of the self-gravitation of the particle are discarded.
In this way, the method uses the WKB approximation in the tunneling formalism for the computation of the imaginary part of the action. The authors Parikh and Wilczek~\cite{Parikh} using the method of radial null geodesic determined the Hawking temperature and in~\cite{Jiang} this method was used by the authors for calculating the Hawking temperature for different spacetimes. In Ref.~\cite{Banerjee} has been analyzed Hawking radiation considering  self-gravitation and back reaction effects in tunneling formalism. It has also been investigated in~\cite{Silva12} the back reaction effects for self-dual black hole using the tunneling formalism by Hamilton-Jacobi method. 
In~\cite{Majumder:afa} has been studied the effects of the GUP in the tunneling formalism for Hawking radiation to evaluate the quantum-corrected Hawking temperature and entropy of a Schwarzschild black hole.
Moreover, the authors in~\cite{Becar:2010zza} have discussed the Hawking radiation for acoustic black hole using  tunneling formalism.

In the literature there are several works on the statistical origin of black hole entropy  --- see for instance ~\cite{Wilczek, Magan:2014dwa,Solodukhin:2011gn}. 
In Ref.~\cite{Kaul}, Kaul and Majumdar computed the lowest order
corrections to the Bekenstein-Hawking entropy. They find that the leading correction is
logarithmic, and is of the type, $\ln\Big({A}/{4}\Big)$ where $ A=4\pi r_h^2 $ is the area of the black hole.
On the other hand, in Ref.~\cite{Carlip:2000nv} was shown that there is an additional logarithmic corrections that depend on conserved charges, i.e.
$\ln[F(Q)]$
where $F(Q)$ is some function of angular momentum and other conserved charges.
In addition, for an understanding of the origin of black hole entropy, 
the brick-wall method proposed by G. 't Hooft has been used for calculations on black holes. Thus, According to G. 't Hooft, black hole entropy is just the entropy of quantum fields outside the black hole horizon. However, when one calculates the black hole statistical entropy by this method, to avoid the divergence of states density near black hole horizon, an ultraviolet cut-off must be introduced. 
In Ref.~\cite{Rinaldi}  was investigated (1 + 1)-dimensional acoustic black hole entropy by the brick-wall method. 
The other related idea in order to cure the divergences is to consider models in which the Heisenberg uncertainty relation is modified, for example in one dimensional space, we have 
\[
\Delta x\Delta p\geq \frac{\hbar}{2}\left( 1 +\alpha^2 (\Delta p)^2 \right) ,  
\]
which shows that there exists a minimal length $ \Delta x\geq \hbar\alpha $, where $ \Delta x $ and
$ \Delta p $ are uncertainties for positon and momentum, respectively, and $ \alpha $ is a positive constant
which is independent of $ \Delta x $ and $ \Delta p $. A commutation relation for the  generalized uncertainty principle (GUP) can be written as 
$[x,p]_{GUP}=i\hbar(1+\alpha^2 p^2)  $, where $ x $ and $ p $ are the positon and the momentum operators, respectively.
Thus, using the modified Heisenberg uncertainty relation the divergence in the brick-wall model are eliminated as discussed in~\cite{Brustein}.
The statistical entropy of various black holes has also been calculated via corrected state density of the GUP~\cite{XLi}. 
Thus, the results show that near the horizon quantum state density and its statistical entropy are finite. In~\cite{KN} a relation for the corrected states density by GUP has been proposed.
The authors in~\cite{ABPS} using a new equation of state density due to GUP~\cite{Zhao}, the statistical entropy of a 2+1-dimensional rotating acoustic black hole has been analyzed. It was shown that considering the effect due to GUP on the equation of state density,  no cut-off is needed \cite{Zhang} and the divergence in the brick-wall model disappears.

In this paper, inspired by all of these previous work we shall focus on the Hamilton-Jacobi method to determine the entropy of a self-dual black hole using the GUP and considering the WKB approximation in the tunneling formalism to calculate the imaginary part of the action in order to determine the Hawking temperature and entropy for self-dual black holes. We anticipate that we have obtained the Bekenstein-Hawking entropy of self-dual black holes and its quantum corrections that are logarithm and also of several other types. 

\section{Tunneling Formalism for self-dual black holes}
The quantum gravitationally corrected Schwarzschild metric of the LBH is
the self-dual black hole metric given by
\begin{equation}
ds^{2} = - F(r)dt^{2} + N(r)^{-1}dr^{2} + H(r)(d\theta^{2} + \sin^{2}\theta\phi^{2}),
\label{self-dual-metric}
\end{equation}
being the metric functions as follows
\begin{eqnarray}
F(r) &=& \frac{(r-r_{+})(r-r_{-})(r+r_{*})^2}{r^{4}+a_{0}^{2}} ,
\\
N(r) &=& \frac{(r-r_{+})(r-r_{-})r^{4}}{(r+r_{*})^{2}(r^{4}+a_{0}^{2})} ,
\\
 H(r) &=& r^{2} + \frac{a_{0}^{2}}{r^{2}},
\end{eqnarray}
where
$r_{+} = {2m}$ is an event horizon,  $r_{-} = 2m P^{2}$ is a Cauchy horizon
and 
$r_{*} = \sqrt{r_{+}r_{-}} = 2mP$,
with $ P \; (P\ll 1)$ the polymeric function given by
\begin{equation}
P = \frac{\sqrt{1+\epsilon^{2}} - 1}{\sqrt{1+\epsilon^{2}} +1} , 
\end{equation}
and $ a_{0} = A_{min}/8\pi $, with $ A_{min} $ being the minimal value of area in Loop Quantum Gravity.
{Note that from the metric (\ref{self-dual-metric}) the suitable 
radial coordinate which is obtained from the function $H(r)$ is
\begin{eqnarray}
R=\sqrt{r^2+\frac{a_0^2}{r^2}}.
\end{eqnarray}
Moreover, the value of $R$ associated with the event horizon is given by
\begin{eqnarray}
R_{h}=\sqrt{H(r_+)}=\sqrt{r_+^2+\frac{a_0^2}{r_+^2}}.
\end{eqnarray}
}
In this section, we will use the tunneling formalism to derive the Hawking temperature for a black hole described by the metric \eq{self-dual-metric}.
In our calculations we assume that the classical action satisfies the relativistic Hamilton-Jacobi equation to leading order in the energy.
Near the black hole horizon the theory is dimensionally reduced to a $2$-dimensional theory \cite{Iso:2006ut,Umetsu:2009ra} whose metric is just the $(t - r)$
sector of the original metric while the angular part is red-shifted away. 
Therefore, the near-horizon metric has the form
\begin{equation}
ds^{2} = - F(r)dt^{2} + N(r)^{-1}dr^{2} \label{r-metric} .
\end{equation}
Now, we consider the Klein-Gordon equations
\begin{equation}
\hbar^{2}g^{\mu\nu}\nabla_{\mu}\nabla_{\nu}\phi - m^{2}\phi = 0,
\end{equation}
under the metric given by \eq{r-metric}
\begin{equation}
- \partial_{t}^{2}\phi + \Lambda \partial_{r}^{2}\phi + \frac{1}{2}\Lambda'\partial_{r}\phi - \frac{m^{2}}{\hbar^{2}}F\phi = 0,
\end{equation}
where $\Lambda = F(r)N(r)$.
{ In this way, using the WKB approximation, we can write}
\begin{equation}
\phi(r, t)=\exp\left[{-\frac{i}{\hbar}{\cal I}(r, t)}\right]. \label{wkb}
\end{equation}
{Then, to the lowest order in $\hbar$, we have}
\begin{equation}
(\partial_{t}{\cal I})^{2} - \Lambda (\partial_{r}{\cal I})^{2} - m^{2} = 0.
\end{equation}
{Now due to the symmetries of the metric, we can suppose a solution of the form}
\begin{equation}
{\cal I}(r,t) = -\omega t + W(r),
\end{equation}
{where for $W(r)$ we have}
\begin{equation}
W = \int \frac{dr}{\sqrt{F(r)N(r)}} \sqrt{\omega^{2} - m^{2}F},
\end{equation} 
In this point, by taking the near horizon approximation 
\begin{eqnarray}
&&F(r) = F(r_+)+F^{\prime}(r_{+})(r-r_{+}) + ... 
\\
&&N(r) = N(r_+)+N^{\prime}(r_{+})(r-r_{+}) + ... ,
\end{eqnarray}
we find that the spatial part of the action function reads
\begin{eqnarray}
W& =& \int \frac{dr}{[F^{\prime}(r_{+})N^{\prime}(r_{+})]^{1/2}}\frac{ \sqrt{\omega^{2} - m^{2}F^{\prime}(r_{+})(r-r_{+})}}{(r-r_{+})},
\nonumber\\
&=&\frac{2\pi i\omega}{[F^{\prime}(r_{+})N^{\prime}(r_{+})]^{1/2}}.
\end{eqnarray}

Moreover, the tunneling probability for a particle with energy $\omega$ is given by
\begin{equation}
\label{gamma}
\Gamma \simeq \exp[-2 Im {\cal I} ] =\exp\Big\{- \frac{\pi\omega[(2m)^{4} + a_{0}^{2}]}{m^{3}(1-P^{2})}\Big\}.
\end{equation}
{Thus, comparing (\ref{gamma}) with the Boltzmann factor 
($ e^{-\omega/T} $), we obtain the Hawing temperature for the self-dual black hole
}
\begin{equation}
T_{sdbh} = \frac{\omega}{Im{\cal I}} = \frac{(2m)^{3}(1-P^{2})}{4\pi[(2m)^{4} +a_{0}^{2}]}.
\end{equation}
{Now, using the laws of black hole thermodynamics the entropy for a self-dual black hole is given by~\cite{l.modesto-prd80, e.alesci-grqc-11015792, s.hossenfelder-grqc12020412}

\begin{equation}
S_{sdbh} =(1+P)^2\int\frac{dm}{T_{sdbh}}= \frac{4\pi(1+P)^{2}}{(1-P^{2})}\Big[\frac{16m^{4} - a_{0}^{2}}{16m^{2}}\Big].
\end{equation}
We can also write the entropy in terms of the area, {of the event horizon area of the self-dual black hole 
($A_{sdbh}=4\pi R^2_{EH}=A+{A^2_{min}}/{(32A)}$ where $  A = 4 \pi r_+ ^ 2=16\pi m^2$ and $A_{min}=8\pi {a_0}$) as follows}
\begin{equation}
S_{sdbh} = \frac{(1+P)^{2}}{(1-P^{2})}\Big[\frac{A}{4}-\frac{\pi^2 a_0^2}{A/4}\Big]
=\frac{(1+P)}{(1-P)}\Big[\frac{A_{sdbh}}{4}-\frac{A^2_{min}}{8A}\Big].
\end{equation}
Therefore, we derive the self-dual black hole temperature and entropy
using the Hamilton-Jacobi version of the tunneling formalism. In particular, we have obtained  for the entropy a quantum correction for the area law. Note that taking the limits, $a_0 \rightarrow 0$ and $P\rightarrow 0$  we have the entropy of a black hole $S = A/4$.
}

\section{ Logarithmic correction to the entropy}
In this section we consider the generalized uncertainty principle (GUP) in the
tunneling formalism and apply the Hamilton-Jacobi method to determine the quantum-corrected Hawking temperature and entropy for a self-dual black hole.
Thus, let us start with the GUP~\cite{ADV, Tawfik:2014zca}
\begin{eqnarray}
\label{gup}
\Delta x\Delta p\geq \hbar\left( 1-\frac{\alpha l_p}{\hbar} \Delta p +\frac{\alpha^2 l^2_p}{\hbar^2} (\Delta p)^2 \right),
\end{eqnarray}
where $\alpha$ is a dimensionless positive parameter, $ l_p=\sqrt{\hbar G/c^3}={M_p G}/{c^2}\approx 10^{-35}m$ is the Planck length, $ M_p =\sqrt{\hbar c/G}$ is the Planck mass and $ c $ is the velocity of light. 
Since that $ G $ is the Newtonian coupling constant,  the correction terms in the uncertainty relation (\ref{gup}) are due to the effects of gravity.

Now the equation (\ref{gup}) can be written as follows.
\begin{eqnarray}
\Delta p\geq \frac{\hbar(\Delta x +\alpha l_p)}{2\alpha^2 l_p^2}
\left(1- \sqrt{1-\frac{4\alpha^2 l_p^2}{(\Delta x +\alpha l_p)^2}}\right),
\end{eqnarray}
where we have chosen the negative sign. Since $ l_p/\Delta x $ is relatively small compared to unity, 
we can expand in Taylor series the equation above
\begin{eqnarray}
\label{p}
\Delta p\geq \frac{1}{2\Delta x}\left[1-\frac{\alpha}{2\Delta x}+ \frac{\alpha^2}{2(\Delta x)^2}+\cdots    \right].
\end{eqnarray}
As we have chosen $ G=c=k_B=1 $, so we also choose $ \hbar=1 $, and we have $ l_p=1 $. In these units 
the generalized uncertainty principle becomes
\begin{eqnarray}
\Delta x\Delta p\geq 1 .
\end{eqnarray}
Now using the saturated form of the uncertainty principle
\begin{eqnarray}
\omega\Delta x\geq1,
\end{eqnarray}
which follows from the saturated form of the Heisenberg uncertainty principle, $ \Delta x\Delta p\geq 1 $, where $ \omega $ is the energy of a quantum particle, then equation (\ref{p}) becomes
 \begin{eqnarray}
\omega_G\geq \omega\left[1-\frac{\alpha}{2(\Delta x)}+ \frac{\alpha^2}{2(\Delta x)^2}+\cdots    \right].
\end{eqnarray}
So if we consider the tunneling formalism via Hamilton-Jacobi method the tunneling probability for a particle with energy corrected $ \omega_G $ reads
\begin{eqnarray}
\Gamma\simeq \exp[-2Im {\cal I}]=\exp\Big[-\frac{2{\pi}  \omega_G}{a}\Big].
\end{eqnarray}
Now comparing with the Boltzmann factor ($ e^{-\omega/T} $), we obtain self-dual black hole temperature 
\begin{eqnarray}
T_{GH}={T}_{sdbh}\left[ 1-\frac{\alpha}{2(\Delta x)}+ \frac{\alpha^2}{2(\Delta x)^2}+\cdots   \right]^{-1}.
\end{eqnarray}
{Here we will choose $ \Delta x=2{R}_{h} $. }
Thus, we have the corrected temperature due to the GUP
\begin{eqnarray}
T_{GH}&=&T_{sdbh}\left[ 1-\frac{\alpha}{4{R}_{h} }+ \frac{\alpha^2}{8{R}_{h}^2 }+\cdots\right]^{-1}
\nonumber\\
&=&\frac{(2m)^{3}(1-P^{2})}{4\pi[(2m)^{4} +a_{0}^{2}]}\left[ 1-\frac{\alpha}{8m}+\frac{a^2_0\alpha}{256m^5}+ \frac{\alpha^2}{32m^2}-\frac{a^2_0\alpha^2}{512m^6}+\cdots\right]^{-1},
\end{eqnarray}
{in powers of $\alpha$ and $a_0$.}

Then, by using the law of black hole thermodynamics we get the entropy
\begin{eqnarray}
S_{G}=(1+P)^2\int\frac{dm}{T_{GH}}=\frac{4\pi(1+P)^2}{128 (1-p^2)}
\left[\frac{\alpha a_0^2}{3m^3}-\frac{8 a_0^2}{m^2}+128 m^2+8 \alpha^2\ln(m)-32\alpha m +O(a^4_0)\right],
\end{eqnarray}
that in terms of $ A=16\pi m^2 $, we have
\begin{eqnarray}
S_{G}&=& S_{sdbh}
+\frac{4\pi(1+P)^2}{128 (1-p^2)}\left[4 \alpha^2\ln\left(S\right)-\frac{16\alpha}{\sqrt{\pi}}\sqrt{S} 
-4 \alpha^2\ln(4\pi)+\frac{(4\pi)^{3/2} \alpha a_0^2}{3S^{3/2}}\right],
\end{eqnarray}
where 
\begin{equation}
S_{sdbh} = \frac{(1+P)^{2}}{(1-P^{2})}\Big[\frac{A}{4}-\frac{\pi^2 a_0^2}{A/4}\Big]
=\frac{(1+P)^{2}}{(1-P^{2})}\Big[S-\frac{\pi^2 a_0^2}{S}\Big],
\end{equation}
is the entropy of a self-dual black hole and $ S=A/4 $.
Note that the coefficient of $ \ln(S)=\ln(A/4) $ is also positive 
and we also get the additional terms $ \sqrt{S}=\sqrt{A/4} $ and $ S^{-3/2}=(A/4)^{-3/2} $ for the entropy.

The black hole energy density can be computed as follows~\cite{Nozari:2008}
\begin{eqnarray}
\frac{8\pi}{3}\rho=-\pi\int S^{\prime}(A)\Big(\frac{A}{4} \Big)^{-2}dA,
\end{eqnarray}
where $ S^{\prime}(A)=\frac{dS(A)}{dA} $. Thus, for the self-dual black hole we have
\begin{eqnarray}
\rho_{sdbh}=\frac{(1+P)^2}{(1-p^2)}\left[\frac{3}{2A}+\frac{8\pi^2a_0^2}{A^3}\right]
=\frac{(1+P)^2}{(1-p^2)}\left[\rho+\frac{64\pi^2a_0^2}{27}\rho^3\right].
\end{eqnarray}
Now the modified energy density for  the self-dual black hole due to the GUP is
\begin{eqnarray}
\rho_{G}=\rho_{sdbh}+\frac{4\pi(1+P)^2}{128 (1-p^2)}\left[\frac{16 \alpha^2}{3}\rho^2
-\frac{24\sqrt{2}\alpha}{\sqrt{3\pi}}\rho\sqrt{\rho}
-\frac{(3072)\sqrt{2}\pi^{3/2}\alpha a^2_0}{189\sqrt{3}}\rho^3\sqrt{\rho}\right].
\end{eqnarray}
The correction of the specific heat capacity  at constant volume $C_v$, reads
\begin{eqnarray}
C_{vG}&=&12\pi T_{GH}=12\pi T_{sdbh}\left[1+\pi\alpha T-{128\pi^5a^2_0\alpha}T^5-\frac{(8\pi)^2\alpha^2}{16}T^2  
+512\pi^6 a^2_0\alpha^2T^6\right]
\nonumber\\
&=&C_{v(sdbh)}\left[1+\alpha\left(\frac{C_v }{12}-\frac{a^2_0 C_v^5 }{1944}\right)
-\alpha^2\left(\frac{4C_v^2}{9}-\frac{a^2_0 C_v^6}{5832}\right) \right],
\end{eqnarray}
where $ T=(8\pi m)^{-1} $ is related to
\begin{eqnarray}
T_{sdbh}=\frac{(1-P^2)T}{(1+2\pi(8\pi)^3 a_0^2T^4)},
\end{eqnarray}
which is the Hawing temperature for the self-dual black hole and then
\begin{eqnarray}
C_{v(sdbh)}=\frac{(12\pi)^4(1-P^2)C_v}{(12\pi)^4+2\pi(8\pi)^3 a_0^2C_v^4},
\end{eqnarray}
is the specific heat capacity for the self-dual black hole.

On the other hand, one can generalize the GUP by considering $ (\Delta x)^2 $ and $(\Delta p)^2$
in the uncertainty relation. Thus, the  symmetric generalized uncertainty principle (SGUP) becomes~\cite{Kim:2007hf,Kempf}
\begin{eqnarray}
\label{sgup}
\Delta x\Delta p\geq \left( 1+\frac{ (\Delta x)^2}{\beta^2} +\alpha^2  (\Delta p)^2 \right),
\end{eqnarray}
Now we solve the equation above for the momentum uncertainty in terms of the position uncertainty 
\begin{eqnarray}
\label{dxdp}
\Delta p\geq \frac{\Delta x}{2\alpha^2}\left[ 1\pm \sqrt{1-\frac{4\alpha^2}{\beta^2}-\frac{4\alpha^2}{(\Delta x)^2}} \right].
\end{eqnarray}
Thus, we can expand the equation (\ref{dxdp}) in Taylor  series as follows 
\begin{eqnarray}
\label{dp}
\Delta p\geq \frac{1}{2\Delta x}\left[1+\frac{(2\Delta x)^2}{\beta^2}
+ \alpha^2\left(\frac{2}{\beta^2}+\frac{(2\Delta x)^2}{\beta^2}+\frac{1}{(2\Delta x)^2}\right)+\cdots    \right].
\end{eqnarray}
Therefore, for equation (\ref{dp}) we can write
 \begin{eqnarray}
\omega_{SG}\geq \omega\left[1+\frac{(2\Delta x)^2}{\beta^2}
+ \alpha^2\left(\frac{2}{\beta^2}+\frac{(2\Delta x)^2}{\beta^2}+\frac{1}{(2\Delta x)^2}\right)+\cdots    \right].
\end{eqnarray}
Using the tunneling formalism via Hamilton-Jacobi method, we obtain the corrected temperature due to the SGUP
\begin{eqnarray}
T_{SGH}&=&T_{sdbh}\left[1+\frac{16R_{h}^2}{\beta^2}
+ \alpha^2\left(\frac{2}{\beta^2}+\frac{16R_{h}^2}{\beta^2}+\frac{1}{16R_{h}^2}\right)+\cdots    \right]^{-1}
\nonumber\\
&=&T_{sdbh}\left[1+\frac{64m^2}{\beta^2} +\frac{4a^2_0}{\beta^2 m^2}
+ \alpha^2\left(\frac{2}{\beta^2}+\frac{64m^2}{\beta^2}+\frac{4a^2_0}{\beta^2 m^2}+\frac{1}{64m^2}
-\frac{a^2_0}{1024 m^6}\right)+\cdots    \right]^{-1}.
\end{eqnarray}
We are now able to obtain the entropy of self-dual black hole 
\begin{eqnarray}
S_{SG}&=&(1+P)^2\int\frac{dm}{T_{SGH}}
=\left(1+\frac{2\alpha^2}{\beta^2}\right)S_{sdbh}
+\frac{4\pi(1+P)^2}{(1-P^2)}\left[\frac{a^2_0}{2048 m^4}-\frac{a^4_0}{8\beta^2 m^4}+\frac{8a^2_0}{\beta^2}\ln(m)\right]
\nonumber\\
&+&\frac{4\pi(1+P)^2}{512\beta^2(1-P^2)}\left[16384(1+\alpha^2)m^4 -\frac{\alpha^2\beta^2a_0^2}{4m^4} +16\Big(256a_0^2(1+\alpha^2)+\alpha^2\beta^2\Big)\ln(m) \right],
\end{eqnarray}
which in terms of horizon area of the self-dual black hole we have
\begin{eqnarray}
\label{entropy-sg}
S_{SG}&=&\left(1+\frac{2\alpha^2}{\beta^2}\right)S_{sdbh}
+\frac{4\pi(1+P)^2}{512(1-P^2)}\left[\frac{1024(1+\alpha^2)}{\pi^2\beta^2}S^2 -\frac{4\pi^2\alpha^2a_0^2}{S^2} +\frac{8}{\beta^2}\Big(256a_0^2(1+\alpha^2)+\alpha^2\beta^2\Big)\ln(S) \right.
\nonumber\\
&-&\left.\frac{8}{\beta^2}\Big(256a_0^2(1+\alpha^2)+\alpha^2\beta^2\Big)\ln(4\pi) \right]
+\frac{4\pi(1+P)^2}{(1-P^2)}\left[\frac{a^2_0\pi^2}{128 S^2}-\frac{2 a^4_0}{\beta^2 S^2}
+\frac{4a^2_0}{\beta^2}\ln(S)+\frac{4a^2_0}{\beta^2}\ln(4\pi)\right].
\end{eqnarray}
Note that in the limit $ \alpha \rightarrow 0 $ of the equation (\ref{entropy-sg}), we obtain
\begin{eqnarray}
\label{ent-gup}
S_{SG}= S_{sdbh}+\frac{4\pi(1+P)^2}{(1-P^2)}\left[\frac{2}{\pi^2\beta^2}S^2  
+\frac{a^2_0\pi^2}{128 S^2}-\frac{2 a^4_0}{\beta^2 S^2}+\frac{8a_0^2}{\beta^2}\ln(S)-\frac{8a_0^2}{\beta^2}\ln(4\pi) \right].
\end{eqnarray}
On the other hand, in the limit $ \beta\rightarrow \infty $ of the equation (\ref{entropy-sg}), we have
\begin{eqnarray}
\label{ent-sgup}
S_{SG}= S_{sdbh} +\frac{4\pi(1+P)^2}{512(1-P^2)}\left[\frac{4\pi^2a^2_0}{S^2}-\frac{4\pi^2\alpha^2a_0^2}{S^2} 
+8\alpha^2\ln(S) -8\alpha^2\ln(4\pi) \right].
\end{eqnarray}
{Therefore, considering both limits we have obtained logarithmic corrections to the entropy of the self-dual black hole. On the other hand,
if $ a_0=0 $, only equation (\ref{ent-sgup}) contains the term $\ln (S)$. Thus, the correction $\ln (S)$ in equation (\ref{ent-gup}) is obtained only if $a_0$ is not zero.
}

\section{Conclusions}
In summary, by considering the GUP, we derive the self-dual black hole temperature and entropy
using the Hamilton-Jacobi version of the tunneling formalism. Moreover, in our calculations the Hamilton-Jacobi
method was applied to calculate the imaginary part of the action and the GUP was introduced by the correction to
the energy of a particle due to gravity near the horizon. The GUP allows us to find quantum corrections to the area
law. Thus, in our model by considering linear  and quadratic GUPs we have obtained logarithmic corrections  
and also the additional terms $ \sqrt{S}=\sqrt{A/4} $ and $ S^{-3/2}=(A/4)^{-3/2} $ for the entropy. The unusual positivity of 
the coefficients of the logarithm terms gives a stronger contribution to the entropy.

\acknowledgments
We would like to thank C.A.S. Silva for useful discussions and CNPq, CAPES, PNPD/PROCAD -
CAPES for partial financial support.

\end{document}